\documentstyle[12pt]{article}

\begin{document}

\title{Pressure induced Raman and fluorescence singularities in $LiYF_4$}

\author{Qiuping A. Wang \\ Institut Sup\'erieur des Mat\'eriaux du Mans, \\
44, Avenue F.A. Bartholdi, 72000 Le Mans, France
\and Alain Bulou and Jean-Yves Gesland \\ Laboratoire de Physique de l'\'Etat Condensé, CNRS UMR6087, \\
Université du Maine, 72085 Le Mans Cedex 9, France}

\date{}

\maketitle

\begin{abstract}
The pressure effect on the fluoride scheelite laser host $LiYF_4$ is studied at room temperature up to 26 GPa
by Raman scattering and up to 40 GPa by $P^{3+}$ fluorescence of doped sample. The Raman spectra exhibit three
singularities at the vicinity of 6 GPa, 10-12 GPa and 16-17 GPa. The samples pressurized to 21 GPa or higher do
not recover the original phase after being released, giving more Raman lines than original samples. The
luminescence spectra of $P^{3+}$ are collected in the energy range corresponding to following transitions
$^3P_{0,1}--^3H_{4,5,6}$, $^1D_2--^3H_4$ and $^3P_0--^3F_2$. Singularities are observed in the vicinity of 6
GPa, 10 GPa, 16 GPa, 23 GPa in agreement with the Raman study. Moreover, an irreversible transition occurs at
23 GPa. The samples pressurized to above 26 GPa become amorphous when released and all the sharp lines
disappear. Above 31 GPa, the spectra at high pressures show only some broad bands corresponding to transitions
between two multiplets of the $^4F_2$ configuration of $Pr^{3+}$. These singularities suggest possible phase
transformations leading to lowering of the lattice symmetry.
\end{abstract}


\newpage

\section{Introduction}
$LiYF_4$ doped with rare earth ions such as $Nd^{3+}$ and $Pr^{3+}$ is a well known laser material with the
tetragonal scheelite structure (Figure 1 and 2) which is known to be one of the rare crystal families
undergoing ferroelastic phase transition with temperature and pressure\cite{Bulo1,Wada,Pinc1,Pinc2,Davi,Benu}.
At ambient condition, this crystal has $C_{4h}^6$ space group. Recently, the lattice dynamics of $LiYF_4$ has
been well studied theoretically and experimentally with IR absorption, Raman and neutron
scattering\cite{Sophie,Mill,Forn1,Sen}. The crystal structure of $LiYF_4$ is relatively stable in temperature,
no transition being observed from 10 K to 1000 K\cite{Sophie,Saran}. But phase transformations under high
pressures have been shown to be possible by Raman and luminescence studies\cite{Saran,Liu}. These works,
especially those under high pressures, are very helpful for understanding the dynamics of this lattice and for
the verification of the validity of the rigid ion models employed in the dynamical calculations.

In a Raman study under pressure up to 20 GPa\cite{Saran}, a sudden change in the Raman frequency slope
$\frac{\partial \omega}{\partial P}$ was observed around 7 GPa and attributed to the stiffening of $LiF_4$
tetrahedra. No anomaly was noted above this pressure. Due to the ambiguïty of mode assignment for high pressure
Raman frequencies, the result of this work is not unquestionable.

$LiYF_4$ doped with rare earth ions gives often very intense emission\cite{Jaya1}. Each rare earth ion is
coordinated to eight F-ligand ions, getting the site symmetry $D_{2d}$ or $S_4$ (figure 1). It is expected that
the luminescence spectra of rare earth ions are very sensible to the behavior (rotation or deformation) of
$LiF_4$ tetrahedra under pressure. A work on $Eu^{3+}:LiYF_4$ has been carried out up to 15 GPa and led to the
conclusion that a phase transition was possible at 10 GPa due to the splitting of some Stake levels meaning
lowering of site symmetry of $Eu^{3+}$.

In this paper, we present the results of the study of $Pr^{3+}:LiYF_4$ with both Raman and Luminescence.
$Pr^{3+}$ in $LiYF_4$ gives very sharp luminescence lines in the visible spectral region with clear line
assignments\cite{Adam,Jaya1,Este,Renf,Gasp}. It was expected that the study can be carried out at much higher
pressures than in previous works and gives us more information about the lattice behavior under pressure. Our
pressure has reached 40 GPa. Several singularities in Raman and luminescence spectra have been observed. It is
concluded that phase transitions induced by pressure are indeed possible.

\section{Raman study}

\subsection{Experiment and result}
The samples are small single crystals of $LiYF_4$ of about 100 $\mu$m width and 50 $\mu$m thick. The pressure
was generated with a gasketed diamond anvil cell Diacell MK3 with a pair of anvils of type IIb/a diamond (0.16
carat) mounted upon tungsten carbide support plates. The anvil flats are 550 $\mu$m in diameter. The gasket in
Inconel was indented before drilling a hole (pressure chamber) in the centre of the indentation. The cylindrical
pressure chamber is 100 $\mu$m in diameter and less than 100 $\mu$m high. A chip of the sample is put into the
pressure chamber together with a small ruby (for pressure measurement) and the pressure medium consisting of a
4:1 mixture of methanol:ethanol. Pressure calibration is carried out by the shift of the R lines of ruby with
the non-linear pressure scale\cite{Mao}. The Raman spectra are excited by the 514.5 nm line of an argon ion
laser with 130 mW total power measured just before the first anvil and collected through the type IIb anvil with
a Dilor-Z24 triple-monochromator single channel Raman spectrometer. The spectra are recorded under microscope
with different orientations of the sample in the pressure chamber in order to obtain as many Raman lines as
possible with different polarisations (since polarization analysis is difficult in the pressure cell). Above 21
GPa, the Raman signals of the samples submitted to monotonic pressure increase become so weak that they could
not be observed even with a very long photon counting time (20 scans of 2 seconds). However, it was found that
better signals were obtained if we used again the samples released from high pressure lightly above 10 GPa.
Under this condition, Raman signals were observed up to 26 GPa.

High pressure Raman spectra are shown in Figure 3. We have observed more lines at low frequency than the
previous work\cite{Saran}. The 13 expected lines at ambient pressure give rise to 9 bands as expected due to
frequency overlaps\cite{Sophie,Mill,Forn1,Sen}. Frequency splitting occurs under small pressure increase so
that 12 lines can be clearly identified (the missing line overlaps the low frequency of $E_g$ line and does not
give significant signal. Their frequencies are reported in Table 1. The significant variation in the intensity
of some lines in the range of 10-14 GPa mainly results from the orientation change of the samples in the
pressure chamber during the experiments. The pressure dependence of the Raman frequencies is plotted in Figure
4. The frequencies of all modes but one increase with increasing pressure with singularities at roughly 6 GPa,
10 GPa and 17 GPa.

\subsection{Discussion of Raman results}
\subsubsection{Preliminary remarks}
Before analyzing the singularities in details, it is useful to give some information about the structure and
its characteristic parameters. The pressure effect mainly results in a contraction of the crystal, proportional
to the elastic constants $S_{11}$ and $S_{33}$. According to reference\cite{Blan} these constants are 12.8
$GPa^{-1}$ and 7.96 $GPa^{-1}$, respectively, which means that the crystal will undergo greater contraction
along the \textbf{a} parameter than along the \textbf{c} parameter. From the microscopic point of view, the
Li-F and Y-F distance are 1.902 $\AA$ and 2.286 $\AA$, respectively, while the shortest sums of the ionic radii
are respectively 1.88 $\AA$ and 2.30 $\AA$\cite{Shan}. Similarly the shortest F-F distances in $LiF_4$ and
$YF_8$ polyhedra are respectively 2.984 $\AA$ and 2.815 $\AA$, while the sum of the ionic radii is 2.57 $\AA$.
This suggests that the contraction of the crystal lattice will result mainly in the reduction or deformation of
$LiF_4$ polyhedra. Finally, it appears from figure 2 that an increase of the angle of the rotation of the
$LiF_4$ tetrahedra around the tetragonal axis should decrease the parameter \textbf{a}, provided the tetrahedra
undergo a distortion in order not to shorten the Li-F distance. So it is expected that increasing pressure
would preferably result in a decrease of \textbf{a}, a decrease of the Li-F bond lengths and an increase of the
rotation angle of $LiF_4$ tetrahedra around the \textbf{c} axis.

\subsubsection{Singularities at 6 GPa}
As observed in the previous work\cite{Saran}, the first singularity around 6 GPa mainly consist in the
splitting of the Bg/Eg lines in the vicinity of 350 $cm^{-1}$ and 400 $cm^{-1}$ (the splitting becomes 65
$cm^{-1}$ for the former and 17 $cm^{-1}$ for the latter at 17 GPa) and a sudden change in the pressure
dependence of the modes at about 350 $cm^{-1}$ and 460 $cm^{-1}$; the latter corresponds to the stretching of
$LiF_4$ tetrahedron\cite{Mill}. We think that a simple stiffening (without other modifications) of $LiF_4$
tetrahedra due to pressure would not be sufficient to account for the abrupt slope change which suggests some
modifications in the structural characteristics (with or without symmetry change). A plausible hypothesis is a
sudden rotation of $LiF_4$ tetrahedra towards the limit value $\phi=45^o$ due to the contraction of the lattice
along \textbf{a} (cf. Figure 2). Note that the decrease of the rotation angle from $\phi=29^o$ to $\phi=0^o$.
(leading to zircon structure) would require the increase of parameter \textbf{a} and so cannot be considered.
In this case of unchanged symmetry, the line splitting is simply the separation of the overlapped Eg and Bg
modes at ambient pressure, as is observed in the work of the reference\cite{Forn1} for the series $LiXF_4$
(X=Y, Ho, Er, Tm, Yb). Another possible interpretation is that it is associated with a symmetry change so the
observed line splitting would rather be a splitting of the Eg modes. This idea is supported by the fact that a
splitting of the luminescence lines of $Pr^{3+}$ in $LiYF_4$ is also observed in this pressure range (see
following section).

\subsubsection{Singularities at 10 GPa}
A second singularity is observed in the vicinity of 10 GPa. There is discontinuity in the pressure shift of the
frequency of the lines located at about 180 $cm^{-1}$, 300 $cm^{-1}$ and 350 $cm^{-1}$ (see Figure 4). We
should emphasize here that a splitting of the luminescence lines of $Pr_{3+}$ in $LiYF_4$ is also observed in
the vicinity of this pressure. In addition, we have recorded at ambient pressure the Raman spectra of the
samples released from pressure of about 11 GPa. The spectrum measured immediately after the releasing (spectrum
c in Figure 5) is very different from the original one (spectrum a in figure 5). Two strong additional lines
are observed at 123 $cm^{-1}$ and 580 $cm^{-1}$. All the lines are very broad, which suggests a partially
amorphous state. This spectrum disappears after roughly 15 minutes for a crystal of about 10 $\mu$m thick and
50 $\mu$m width which recovers the original Raman lines (spectrum b in Figure 5). Since the structure above 6
GPa is not known, no mechanism can be proposed for this structure transformation. As a matter of fact, any of
the mechanisms proposed to explain the singularities at 6 GPa is possible in the structural change at 10 GPa.

\subsubsection{Singularities at 17 GPa}
A third singularity appears at about 17 GPa. It is mainly characterized by the vanishing of some lines between
400 $cm^{-1}$ and 450 $cm^{-1}$, the emergence of a new line at 435 $cm^{-1}$, a change of the pressure
dependence of the frequency of the Ag mode at about 315 $cm^{-1}$, and the existence of a minimum of the low
frequency Eg mode (at 154 $cm^{-1}$). At higher pressure, the energy of this mode increases again with a larger
shift rate than below 17 GPa and reaches 156 $cm^{-1}$ at 26 GPa. The most striking fact is that the Raman
spectra (spectrum d in figure 5) recorded at ambient pressure of the samples released from 21 GPa remain
definitely different from those of the sample never pressurized. This means that an irreversible transformation
has occurred. Note that the frequencies of the new spectrum (d in figure 5) are close to those of the samples
released from 11 GPa (c of figure 5).

The irreversibility and the significant change of the high pressure Raman spectra suggest that this very high
pressure phase transformation would be reconstructive. However, as previously pointed out, it seems to be
associated with a softening of the low frequency Eg mode, as would occur for a second order (reversible) phase
transition. This can be seen in figure 6 where the square of the frequency is plotted against pressure, giving
following relationship :
\begin{equation}                                        \label{1}
\omega^2=A(p-p_0)
\end{equation}
for $P<17$ GPa, where $A=300$ $cm^{-2}GPa^{-1}$ and $p_0= 80$ GPa; and :
\begin{equation}                                        \label{2}
\omega^2=A'(p_0'-p)
\end{equation}
for higher pressure with $A'=794$ $cm^{-2}GPa^{-1}$ and $p_0'=-7$ GPa.

In this framework, this phase transition would be a proper ferroelastic one since the soft mode, observed below
and above the transition, has the same symmetry as a strain tensor\cite{Bulo1}. It is worth noticing that, due
to its degeneracy, it would be associated with the existence of a soft acoustic plane \cite{Cowl}, a kind of
transition expected but never experimentally observed to our knowledge. We recall that the scheelite $BiVO_4$
undergoes also a proper ferroelastic phase transition in which the softening of an acoustic branch was
observed. But the soft mode has a one dimensional Bg symmetry. The values of A and A' for $LiYF_4$ are about
three times smaller than those for $BiVO_4$, but $(A'/A)=2.61\pm0.5$ for $LiYF_4$ is very close to
$(A'/A)=2.67\pm0.2$ observed for $BiVO_4$ \cite{Pinc2}. What is really questionable is the fact that such a
transition, clearly associated to a soft mode, is irreversible. Actually a similar phenomenon has already been
observed in the layered tetrafluoroaluminate $KAlF_4$ in which an irreversible martensitic phase transition
appears after the softening of a flat Brillouin zone boundary phonon branch \cite{Bulo2}. So, the softening of
the low frequency Eg mode may be associated with a similar mechanism.

\section{Luminescence study}

\subsection{Experimental results}
The technical details of the pressure experiment are described in the above section. The sample is $LiYF_4$
single crystal doped with approximately 1 mol $Pr^{3+}$. The fluorescence spectra are excited at 457.9 nm with
an argon ion laser. The pressure goes up to 40 GPa and the emission spectra are recorded in the range 15000
$cm^{-1}$--21500 $cm^{-1}$ corresponding to the transitions from $^3P_0$, $^3P_1$ and $^1D_2$ initial states to
$^3H_4$, $^3H_5$, $^3H_6$ and $^3F_2$ final states.

The observed lines at ambient pressure are summarized in Table 2. The assignments of the transitions are given
for $S_4$ site symmetry of the $Pr^{3+}$ ion \cite{Adam,Este}. Twenty four emission lines ($L_1$ to $L_{24}$)
are assigned to the transitions between five initial states and fourteen final ones. The $^3H_5$ level at 2673
$cm^{-1}$ was not observed before and is determined in this work by two sharp lines at 18197 $cm^{-1}$ (from
$^3P_0$) and 18752 $cm^{-1}$ (from $^3P_1$). It is assigned to $\Gamma_{3,4}$ symmetry according to the
calculated result of reference \cite{Este}.

The pressure effect on the observed fluorescence spectra is shown in figure 7. In figure 8 is plotted the
pressure shift of the observed emission lines. It can be seen that, in spite of some vanishing lines, the
number of the emission lines increases constantly up to 31 GPa at which all the lines abruptly vanish. The
remaining spectra contain only some broad emission bands showing a weak pressure shift up to 40 GPa. Briefly,
singularities are observed at 6 GPa (26 lines), 10 GPa (32 lines), 16 GPa (34 lines), 23 GPa (37 lines) and 31
GPa (vanishing of all the lines). They will be discussed below in detail.

\subsection{Discussion of fluorescence results}

\subsubsection{Singularity at 6 GPa}
The first singularity occurs at 6 GPa as in Raman spectra. Two strong lines ($L_7$ and $L_8$) from the
transitions $^3P_1$($\Gamma_{3,4}$)-$^3H_5$($\Gamma_{2}$) and $^3P_1$($\Gamma_{3,4}$)-$^3H_5$($\Gamma_{3,4}$)
vanish (figure 8b) and two additional lines at 18590 $cm^{-1}$ ($L_{n1}$) and 15530 $cm^{-1}$ ($L_{n2}$) are
observed. $L_{n1}$ cannot be interpreted by the splitting of the 18592 $cm^{-1}$ line ($L_9$) from the
transition $^3P_0$($\Gamma_{1}$)-$^3H_5$($\Gamma_{2}$) between two singlets. It could be attributed to the
transitions $^3P_0$($\Gamma_{1}$)-$^3H_5$($\Gamma_{3,4}$), situated at 8 $cm^{-1}$ above $L_9$ and observed at
low temperature[9,10]. That means that the appearance of $L_{n1}$ is only a pressure induced separation between
two lines close to each other.

$L_{n2}$ in figure 8d seems to be the result of a splitting of the line $L_{21}$ from the transition
$^3P_0$($\Gamma_{1}$)-$^3F_2$($\Gamma_{3,4}$) which is broadened before splitting, as shown in figure 7b.
However it is difficult to conclude that this comes from the lifting of the degeneracy of the
$^3F_2$($\Gamma_{3,4}$) level in a lower site symmetry, since no splitting of other levels of $\Gamma_{3,4}$
symmetry was observed at this pressure. As a matter of fact, $L_{n2}$ could be related to one of the
transitions $^3P_0$($\Gamma_{1}$)-$^3F_2$($\Gamma_{3,4}$) and $^3P_0$($\Gamma_{1}$)-$^3F_2$($\Gamma_{2}$) with
the latter situated at 20 $cm^{-1}$ above the former and much less intense than the former at ambient
pressure[9]. In this case, the observed splitting of $L_{21}$ could be only the effect of different pressure
dependence of two lines that fortuitously overlap at ambient pressure.

According to the above discussion, it is not sure that the singularity observed at 6 GPa are induced by a phase
transition.

\subsubsection{Singularity at 10 GPa}

The singularity in the Raman spectra observed at this pressure is only a weak variation in the shift rate of
the Raman mode as a function of pressure.  In fluorescence, as shown in figure 8c, following additional lines
are observed : $L_{n3}$ (19140 $cm^{-1}$), $L_{n4}$ (19020 $cm^{-1}$), $L_{n5}$ (17100 $cm^{-1}$), $L_{n6}$
(16400 $cm^{-1}$) and $L_{n7}$ (16300 $cm^{-1}$). A line ($L_{19}$) vanishes. The slope of $L_5$, $L_{12}$,
$L_{13}$, $L_{14}$, $L_{21}$ and $L_{23}$ abruptly changes.

$L_{n4}$ emerges in the neighborhood of a strong line $L_5$ from the transition
$^3P_1$($\Gamma_{3,4}$)-$^3H_5$($\Gamma_{2}$). This looks like the splitting of the $^3P_1$($\Gamma_{3,4}$)
level. But as seen in figure 8a, the lines $L_1$ and $L_2$ of the transitions
$^3P_1$($\Gamma_{3,4}$)-$^3H_4$($\Gamma_{2}$) and $^3P_1$($\Gamma_{3,4}$)-$^3H_4$($\Gamma_{3,4}$) respectively
do not split. So $L_{n3}$ and $L_{n4}$ are probably due to the transitions from $^3P_1$($\Gamma_{3,4}$) to two
of the levels close to $^3H_5$($\Gamma_{2}$) at 2280 $cm^{-1}$, i.e. the levels at 2253 $cm^{-1}$
($\Gamma_{1}$), 2272 $cm^{-1}$ ($\Gamma_{3,4}$) and 2297 $cm^{-1}$ ($\Gamma_{1}$) (values at ambient pressure),
which were not observed at ambient pressure [9].

$L_{n6}$ and $L_{n7}$ are close to $L_{18}$ from the transition $^3P_0$($\Gamma_{1}$)-$^3H_6$($\Gamma_{3,4}$),
but none of them seems to be the result of the splitting of the $^3H_6$($\Gamma_{3,4}$) final state. In
addition, another $^3P_0$($\Gamma_{1}$)-$^3H_6$($\Gamma_{3,4}$) transition ($L_{17}$) remains unchanged. So
more probably $L_{n6}$ and $L_{n7}$ result from the transitions from $^3P_0$($\Gamma_{1}$) to
$^3H_6$($\Gamma_{1}$) (4430 $cm^{-1}$) and $^3H_6$($\Gamma_{1}$) (4486 $cm^{-1}$) in the neighborhood of
$^3H_6$($\Gamma_{3,4}$) at 4456 $cm^{-1}$ at ambient pressure[9]. It should be noted that the transitions from
$^3P_0$($\Gamma_{1}$) to $^3H_6$($\Gamma_{1}$) and $^3H_6$($\Gamma_{1}$) are forbidden by the electric-dipole
selection rules in $S_4$ site symmetry[9,10].

Another singularity at this pressure is related to a sharp line at 19260 $cm^{-1}$ (see figure 8b). It cannot
be attributed to any transition between the electronic or vibronic states[9-12]. It becomes very weak at about
10 GPa and rapidly disappears in the broad band of the $^3P_1$($\Gamma_{3,4}$)-$^3H_5$($\Gamma_{2}$) emission,
as shown in figure 7a.

To conclude, a structure transformation must be invoked to explain the changes in the emission spectra, a
structural modification leading to the change of transition probability and selection rules in the $S_4$ site.
This supports the suggestion of the previous work \cite{Liu} and our Raman result that the ambient pressure
spectrum of the samples released from 11 GPa was different from that of the scheelite structure $LiYF_4$.

\subsubsection{Singularity at 16 GPa}

We recall that a minimum of a pressure induced soft Raman E mode has been observed around this pressure. In
fluorescence, three additional lines $L_{n8}$ (18560 $cm^{-1}$), $L_{n9}$ (18470 $cm^{-1}$) and $L_{n10}$
(15470 $cm^{-1}$) appear and two lines $L_{20}$ and $L_{23}$ vanish.

$L_{n8}$ and $L_{n9}$ appear between $L_9$ and $L_{n1}$, and become two sharp lines up to 30 GPa (figure 7a).
Their pressure shift being almost parallel to those of the neighbouring lines (figure 8b), they could be
integrated into the $^3P_0$-$^3H_5$ line group and attributed to the transitions from $^3P_0$ to two of the
three states $^3H_5$($\Gamma_{1}$), $^3H_5$($\Gamma_{3,4}$) and $^3H_5$($\Gamma_{1}$), situated at ambient
pressure at 2253 $cm^{-1}$, 2272 $cm^{-1}$ and 2297 $cm^{-1}$ respectively[9]. It should be noticed that,
together with $L_{n8}$ and $L_{n9}$, there are now seven observed lines in the $^3P_0$-$^3H_5$ group, which
exceeds the number (five) of the transitions allowed by the electric-dipole selection rules in $S_4$ site
symmetry[9,10].

$L_{n10}$ is in the $^3P_0$-$^3F_2$ line group (figure 8d) and could only be considered as a transition from
$^3P_0$($\Gamma_{1}$) to one of the $^3F_2$ levels resulting from the splitting of the $^3F_2$($\Gamma_{3,4}$)
level, or to the fourth Stark level of $^3F_2$ multiplet situated at 5159 $cm^{-1}$ ($\Gamma_{1}$) to which the
transition from $^3P_0$($\Gamma_{1}$) is forbidden in $S_4$ symmetry at ambient pressure. In any case, the
appearance of $L_{n10}$ suggests a change in the site symmetry of $Pr^{3+}$. According to Raman spectra, this
transition would be a second order one.

\subsubsection{Singularity at 23 GPa}

At 23 GPa, nine additional lines emerge. They are $L_{n11}$ (20709 $cm^{-1}$), $L_{n12}$ (20609 $cm^{-1}$),
$L_{n13}$ (20489 $cm^{-1}$), $L_{n14}$ (18995 $cm^{-1}$), $L_{n15}$ (18596 $cm^{-1}$), $L_{n16}$ (16580
$cm^{-1}$), $L_{n17}$ (16460 $cm^{-1}$), $L_{n18}$ (16425 $cm^{-1}$), $L_{n19}$ (16275 $cm^{-1}$) and
$L_{n20}$(15510 $cm^{-1}$). On the other hand, $L_6$, $L_{13}$, $L_{18}$, $L_{24}$, $L_{n5}$ and $L_{n10}$
vanish (figure 8).

$L_{n11}$, $L_{n12}$ and $L_{n13}$ are in the $^3P_0$-$^3H_4$ group and can only be attributed to the
transitions from $^3P_0$ to three $^3H_4$ levels in the energy range 0-300 $cm^{-1}$ (see figure 8a) just above
the $^3H_4$($\Gamma_{2}$) ground state. This, together with $L_3$ and $L_4$, gives rise to five energy levels
in this region where only four levels should be observed in the $S_4$ site symmetry, namely
$^3H_4$($\Gamma_{2}$), $^3H_4$($\Gamma_{3,4}$) and two $^3H_4$($\Gamma_{1}$)[9]. So it is reasonable to suppose
that the degenerate $^3H_4$($\Gamma_{3,4}$) level splits into two levels giving rise to $L_{n11}$ and the
remaining $L_4$. This splitting should exist also in the $^3P_1$-$^3H_4$ group ($L_1$ and $L_2$) but
unfortunately it is not possible to be observed due to the extreme signal weakness at high pressure.

Above 23 GPa, together with $L_{n15}$, there are eight observed lines in the $^3P_0$-$^3H_5$ (figure 8b) group
and nine in the $^3P_0$-$^3H_6$ group (figure 8c) while only seven lines are allowed for the latter in $S_4$
symmetry[9,10].

Finally, in the $^3P_0$-$^3F_2$ group (figure 8d), together with $L_{n17}$, five lines are observed, exactly
the number of states in this multiplet. This means that all the energy levels are singlet at this pressure.

Another significant phenomenon is that, as in our Raman study, the luminescence of the samples released from
this pressure are different from that of the samples never pressurized (Figure 9).

\subsubsection{Singularity at 31 GPa}

In the pressure range above 23 GPa, the sharp lines still exist up to 31 GPa, a pressure at which all the lines
suddenly disappear (figure 7 and 8). Each of the broad bands corresponds to a transition from $^3P_0$ or
$^3P_1$ to one of the multiplets $^3H_{4-6}$ and $^3F_{2-4}$. The spectra remain the same up to 40 GPa with
only a small pressure red shift of the broad bands (see figure 7 and 8). This means that the lattice of
$LiYF_4$ becomes amorphous above 31 GPa.

Similarly to the samples pressurized to above 23 GPa, the samples released from pressures higher than 31 GPa
remain definitely amorphous at normal pressure, giving only broad luminescence bands as shown in Figure 7.

\section{Final comment}
In summary, we have presented in this paper the experimental results of pressure studies of $LiYF_4$ structure
by both Raman scattering and $Pr^{3+}$ luminescence. The pressure runs up to 40 GPa. This work gives us more
information about the behaviors of the lattice structure of $LiYF_4$ than the previous ones\cite{Saran,Liu}.
The Raman and fluorescence results are in agreement concerning the pressure effect on the structure of
$LiYF_4$. Singularities are observed in the vicinity of 6 GPa, 10 GPa, 16 GPa, 23 GPa and 31 GPa. Most of these
singularities suggest that a simple stiffening of the lattice or the tetrahedra is not sufficient to account
for the pressure evolution of the spectra and that the scheelite type structure may undergo a series of
transformations or distorsions responsable for the lowering of the site symmetry and the singularities at 10
GPa, 16 GPa, 23 GPa and 31 GPa. The transformations at 23 GPa and 31 GPa are irreversible, leading to amorphous
structures. Of course, further work such as high pressure X-ray diffraction is necessary to clarify the
situation.

With regard to the mechanism of the possible structure transformations of $LiYF_4$ under high pressure, no
precision can be given for the time being. But any of the mechanisms including the rotation and displacement of
the $LiF_4$ tetrahedra could be invoked. It is worth noticing that the internal binding of the $LiF_4$
tetrahedra is not entirely different in comparison with the external binding[12,13], i.e. the ionic Li-F and
F-F forces in the tetrahedra are of similar magnitude to the ionic Y-F and F-F forces outside the tetrahedra.
So when there is a lattice distortion, we should not exclude the possibility of the distortion of the
tetrahedra and of the emergence of new kinds of polyhedra under very high pressure.

\newpage

{\huge Table and Figure captions}

Table 1: Observed Raman modes at ambient pressure, their pressure induced frequency shift $\delta\nu_i$ and
Gr\"uneisen parameters $\gamma_i$ up to 17 GPa. $\beta$ is the isothermal compressibility coefficient. L or H
means the lower or higher frequency branch of the split modes above 6 GPa and below 26 GPa.

Table 2: Luminescence lines (in $cm^{-1}$) of $LiYF_4:Pr^{3+}$ at ambient pressure. The assignment of the
transitions is given in the irreducible representations $\Gamma_1$, $\Gamma_2$, $\Gamma_3$ and $\Gamma_4$ of
the $S_4$ point group. The degenerate $\Gamma_3$ and $\Gamma_4$ are denoted by $\Gamma_{3,4}$.

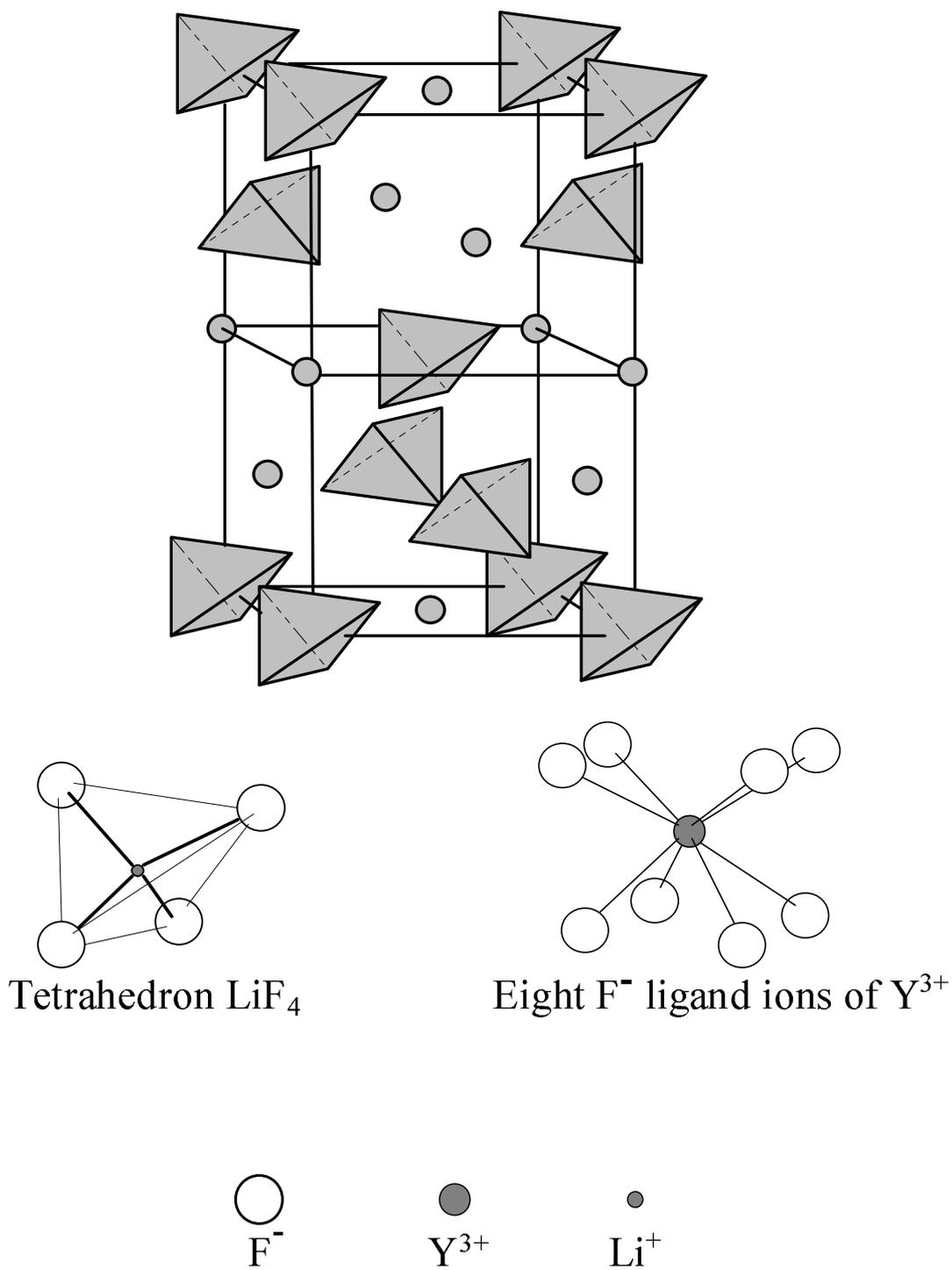
Figure 1, Scheelite structure of $LiYF_4$ with space group  $C_{4h}^6$. $Y^{3+}$ ion is coordinated to eight
$F^-$ ligand ions at the corners of the eight $LiF_4$ tetrahedra around it and has a $S_4$ or $D_{2d}$ site
symmetry.

Figure 2, Projection of the tetrahedra in a unit cell of the scheelite structure of $LiYF_4$ on the (001)
plane. The full circles represent the eight $F^-$ ligands of $Y^{3+}$.

Figure 3, Pressure influence on the Raman spectra of $LiYF_4$. The variation in line intensity of the spectra
below 16 GPa is mainly due to the different orientations of the samples during the experiments.

Figure 4, Pressure dependence of the Raman frequency of $LiYF_4$. The size of the symbols represents the error
of the measurements. The dashed lines are just guides for the eye.

Figure 5, Comparison of the spectra recorded at ambient pressure for $LiYF_4$ after different treatments: (a)
no treatment; (b) sample pressurized to 11 GPa, recorded about 15 minutes after releasing the sample from
pressure cell; (c) sample as in (b) but recorded immediately after releasing the sample; (d) sample released
from 21 GPa. This spectrum remains different from that of (a) or (b).

Figure 6, pressure dependence of the square of the pressure-induced soft phonon energy. The size of the symbols
represents the error of the measurements.

Figure 7: Pressure influence on the luminescence spectra of $Pr^{3+}$ in $LiYF_4$ : (a) $^3P_{0,1}-^3H_5$
transitions, (b) $^3P_{0,1}-^3H_6$ and $^3P_{0,1}-^3F_2$ transitions.

Figure 8: Pressure dependence of the fluorescence lines from the transitions (a) $^3P_{0}-^3H_4$; (b)
$^3P_{0}-^3H_5$; (c) $^3P_{0}-^3H_{4,6}$, $^3P_{0}-^3F_2$ and $^1D_{2}-^3H_{4,6}$; and (d) $^3P_{0}-^3F_{2}$;
for $Pr^{3+}$ in $LiYF_4$. The size of the symbols represents the standard error of the measurements. See table
1 for the assignment of the lines L1 to L4.

Figure 9: Emission spectrum recorded at ambient pressure for $LiYF_4:Pr^{3+}$ released from 23 GPa.

\end{document}